\documentclass[preprint]{aastex}

\usepackage{amsmath}
\shorttitle{POLAR REGION AND QUIET REGION OF THE SUN}
\shortauthors{H. Ito et al.}
\begin{document}

\title{IS THE POLAR REGION DIFFERENT FROM THE QUIET REGION OF THE SUN?}

\author{HIROAKI ITO\altaffilmark{1}, 
SAKU TSUNETA\altaffilmark{2},
DAIKOU SHIOTA\altaffilmark{1},
MUNETOSHI TOKUMARU\altaffilmark{1},
KEN'ICHI FUJIKI\altaffilmark{1} }

\altaffiltext{1}{Solar-Terrestrial Environment Laboratory, 
Nagoya University, Furo-cho, Chikusa-ku, Nagoya, 464-8601, Japan}
\altaffiltext{2}{National Astronomical Observatory of Japan (NAOJ)
Mitaka, Tokyo 181-8588, Japan}

\begin{abstract}
Observations of the polar region of the Sun are critically important for understanding the solar dynamo and the acceleration of solar wind. We carried out  precise magnetic observations on both the North polar region and the quiet Sun at the East limb with the Spectro-Polarimeter of the Solar Optical Telescope aboard {\it Hinode} to characterize the polar region with respect to the quiet Sun. The average area and the total magnetic flux of the kG magnetic concentrations in the polar region appear to be larger than those of the quiet Sun. The magnetic field vectors classified as {\it vertical} in the quiet Sun have symmetric histograms around zero in the strengths, showing balanced positive and negative flux, while the histogram in the North polar region is clearly asymmetric, showing a predominance of the negative polarity. The total magnetic flux of the polar region is larger than that of the quiet Sun. In contrast, the histogram of the {\it horizontal} magnetic fields is exactly the same between the polar region and the quiet Sun. This is consistent with the idea that a local dynamo process is responsible for the horizontal magnetic fields. A high-resolution potential field extrapolation shows that the majority of magnetic field lines from the kG-patches in the polar region are open with a fanning-out structure very low in the atmosphere, while in the quiet Sun, almost all the field lines are closed. 
\end{abstract}


\section{INTRODUCTION}
The polar regions of the Sun have some distinct properties compared with those of the quiet Sun: (1) the coronal holes in the polar regions are extended and stationary, (2) the fast solar wind emanates from the polar region, (3) the meridional flow may reach the polar regions, which must have a sink for the flow coming to the polar regions, and (4) the global poloidal magnetic field of the Sun manifests itself in the polar regions. Therefore, information on the magnetic properties of the polar regions is critically important for the understanding of the solar dynamo problem. The acceleration mechanism of the fast solar wind must also be related to the spatial and temporal structures of magnetic fields on the surface of the Sun, and the physical process creating the wind is still a topic of debate  \citep{Wang89, Wang90, McCo00, Kojima01, Tu05}. 

The polar regions have been observed with ground-based magnetograms \citep{Babcock55, Severny71, svalgaard78, Tang91, Lin94, Homann97, Fox98, Oku04, Blanco07}. These observations as well as observations with {\it SOHO/MDI} \citep{Bene04} have provided measurements of the line-of-sight magnetic component. For more accurate observations, full Stokes polarimetry for the polar regions has been carried out \citep{Lites96}, but because of the strong intensity gradient and the foreshortening effect at the solar limb, the spatial resolution of the ground-based measurements has been compromised. The magnetic properties of the polar region are poorly understood. 

\citet{tsuneta08} found with {\it Hinode} \citep{Kosugi07} that there are patchy concentrations of magnetic fields with field strengths exceeding 1 kG in the polar region, confirming the result of \citet{Oku04}. These magnetic fields are almost {\it vertical} to the local surface of the Sun. Ubiquitous magnetic concentrations with field strengths larger than 1 kG are known to exist in the quiet Sun as well \citep[e.g.,][]{david07a}. This paper focuses on determining whether there is any difference in the magnetic properties between the polar region and the quiet Sun. If there is a difference, what is the implication to the problems of solar dynamo and the acceleration of the fast solar wind? With this motivation, we observed the polar region and the quiet Sun as a fiducial data for the analysis of the polar region with the spectropolarimeter (SP) \citep{Lites01} of the Solar Optical Telescope (SOT) \citep{tsuneta08a, Suematsu08, Ichi08, Shimizu08} aboard {\it Hinode}.  

The sensitivity to a given magnetic field vector depends on its orientation with respect to the observer and on the Stokes parameter used to detect it. Stokes {\it Q} and {\it U} signals are used to detect the field component perpendicular to the line-of-sight, while Stokes {\it V} signal is used for the field component parallel to the line-of-sight. Measurement through Stokes {\it V} is more sensitive than that through Stokes {\it Q} and {\it U} when the field is weak. We observed the quiet Sun located at the East limb instead of the quiet Sun at the disk center, so that the sensitivity to detect any weak magnetic field is the same for both the polar region and the quiet Sun. This allows a direct comparison between the polarimetric data for the polar region with that for the quiet Sun.  

\begin{figure}
 \begin{center}
  \epsscale{.80}
  \plotone{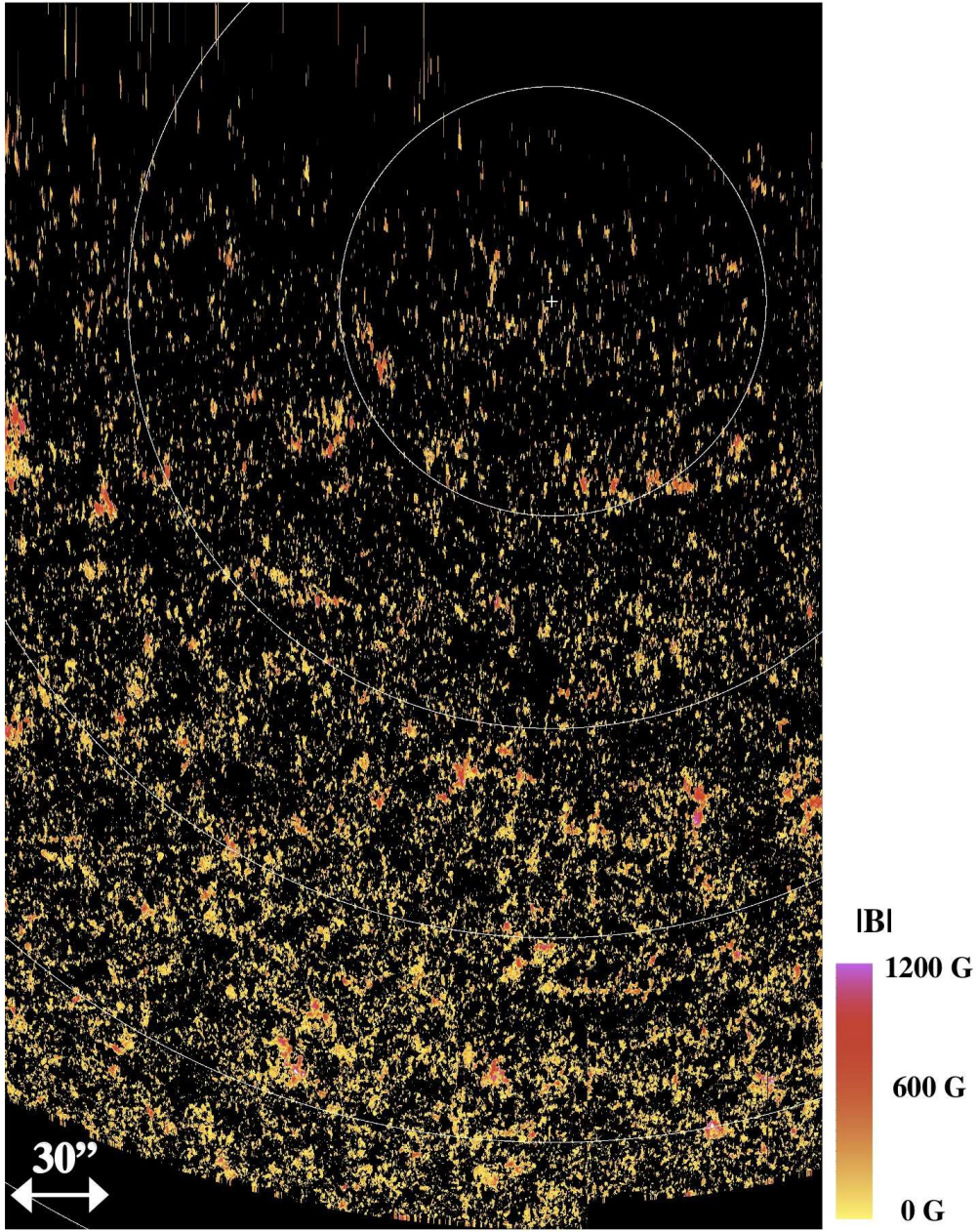}
\end{center}
\caption{
Magnetic landscape of the polar region of the Sun; The map of the magnetic field strength on the sky plane is converted to the map seen from above the North pole. The observations took place at 00:10---07:26 UT on 2007 September 25. East is to the left. The pixel size is $0.16''$. The magnetic field strengths were obtained only for pixels whose polarization signal exceeds a given threshold (see text). The field of view is $320''$ by $477.6''$. The size of the field of view for North-South direction ($163.84''$) is expanded to $477.6''$ as a result of correction for foreshortening. The circular arc is the latitudinal line per $5^\circ$ from the North pole. 
}
\label{npole2}
\end{figure}

\begin{figure}
 \begin{center}
  \epsscale{1.}
  \plotone{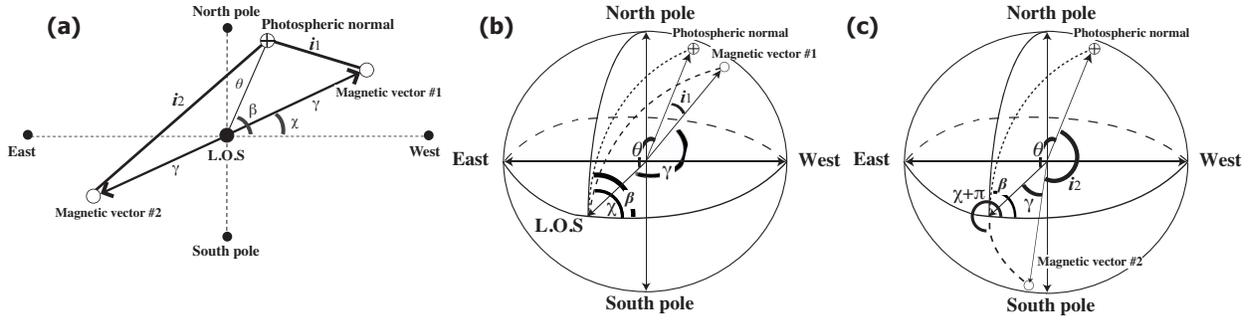}
 \end{center}
 \caption{
Two solutions of the magnetic field vectors $\sharp$1 and $\sharp$2 with the inclination angles of ${\it i_{\rm 1}}$ and ${\it i_{\rm 2}}$ with respect to the photospheric normal for the two possible azimuth angles of $\chi$ and $\chi + \pi$ due to the $180^\circ$ ambiguity, respectively. $\theta$ is the angle between the local normal and the line-of-sight (L.O.S.), $\gamma$ the inclination ($ 0 \leq \gamma \leq \pi $) of the magnetic field vector with respect to L.O.S., and $\chi$ is the azimuth angle ($ 0 \leq \chi \leq \pi $). Panel $(a)$ shows these angles on the sky plane. Note that the lines indicating angles are represented by the straight lines, not by the segments of arc, for simplicity. Panel $(b)$ indicates the same information in terms of vectors at the position of any observing pixel. The dotted and dashed lines represent Great-circles connecting the L.O.S. direction and the photospheric normal and the magnetic vector $\sharp$1, respectively. Panel $(c)$ is the same as the Panel $(b)$ but for the magnetic vector $\sharp$2. These two representations shown in the panel $(a)$, and the panels $(b)$ and $(c)$ are equivalent.
}
 \label{incli1}
\end{figure}

\begin{figure}
 \begin{center}
  \epsscale{1.0}
  \plotone{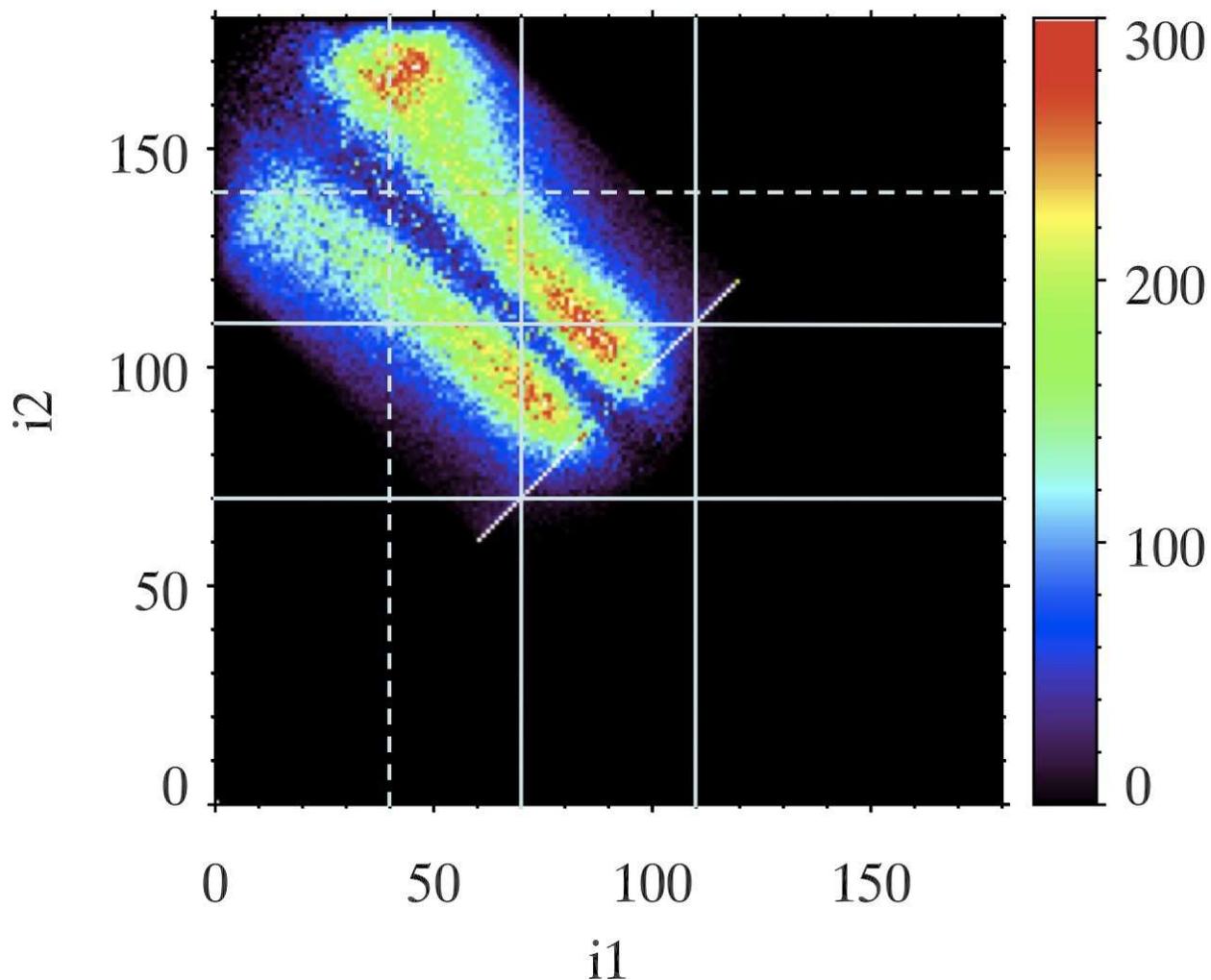}
 \end{center}
 \caption{
Scatter plot of the two solutions ${\it i_{\rm 1}}$ and ${\it i_{\rm 2}}$ of the inclination angles with respect to the local normal to the photosphere for the North pole. The color table indicates the density of the data points per $1^\circ \times 1^\circ $. Magnetic fields are classified as {\it vertical} to the local surface if they are located either close $0^\circ$ or $180^\circ$ (dashed lines), and as {\it horizontal} if located near $90^\circ$ (bold lines). For the concentration along the line ${\it i_{\rm 1}} \sim {\it i_{\rm 2}}$ as indicated by the white line, see text.
}
 \label{incli2}
\end{figure}

\begin{figure}
 \begin{center}
  \epsscale{.80}
  \plotone{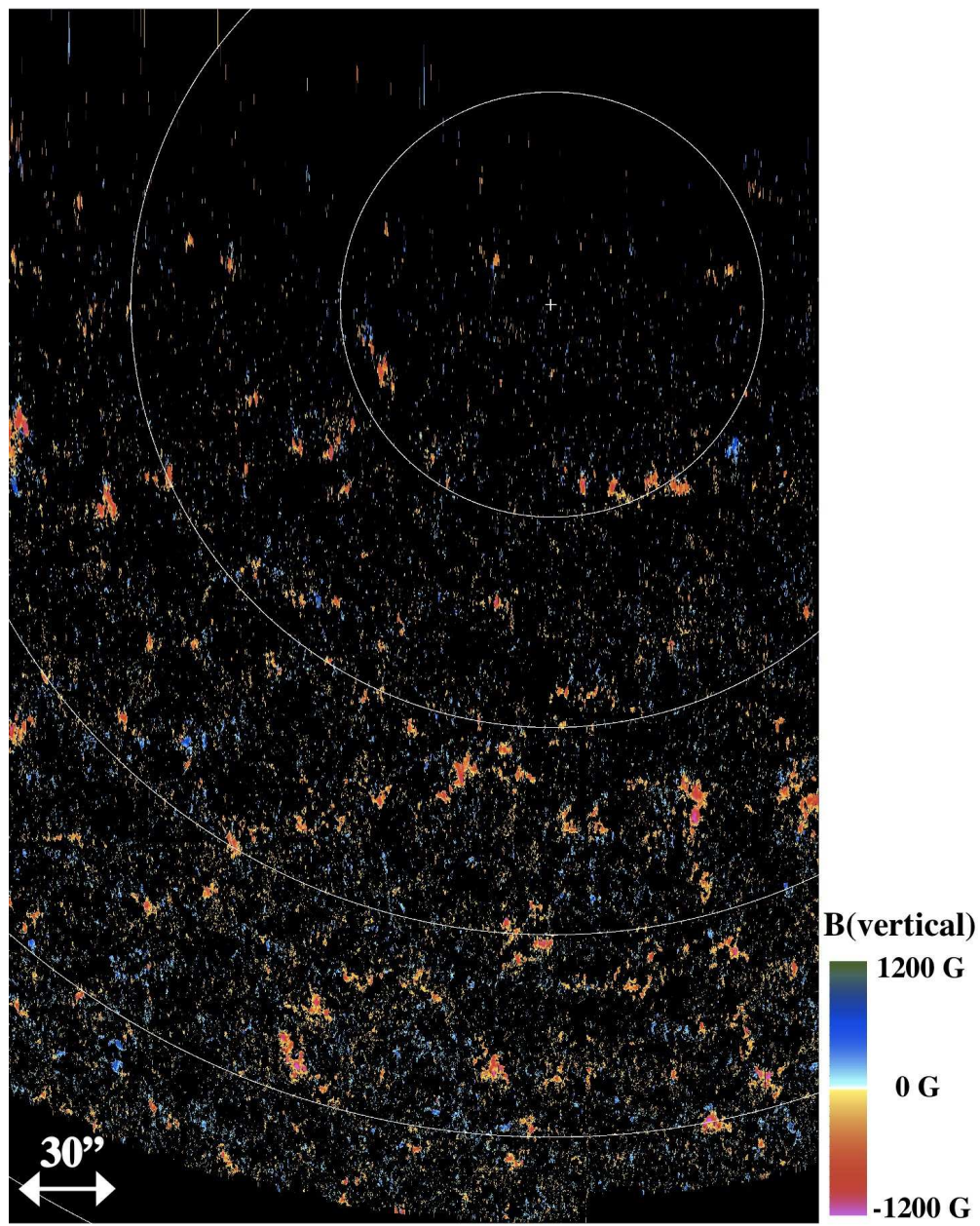}
 \end{center}
 \caption{
{\it (a)} Map of signed strength of the magnetic field vectors classified as {\it vertical} in the polar region. The original data is the same as that of Fig. \ref{npole2}. The pixel size is $0.16''$. The magnetic field strength was obtained only for pixels whose polarization signal exceeds a given threshold (see text). These panels can be directly compared with the quiet Sun maps (Fig. \ref{elimb3}). The scale size and the color table for these maps are the same.  
}
\label{npole3}
\end{figure}
\begin{figure}
 \begin{center}
  \epsscale{.80}
  \plotone{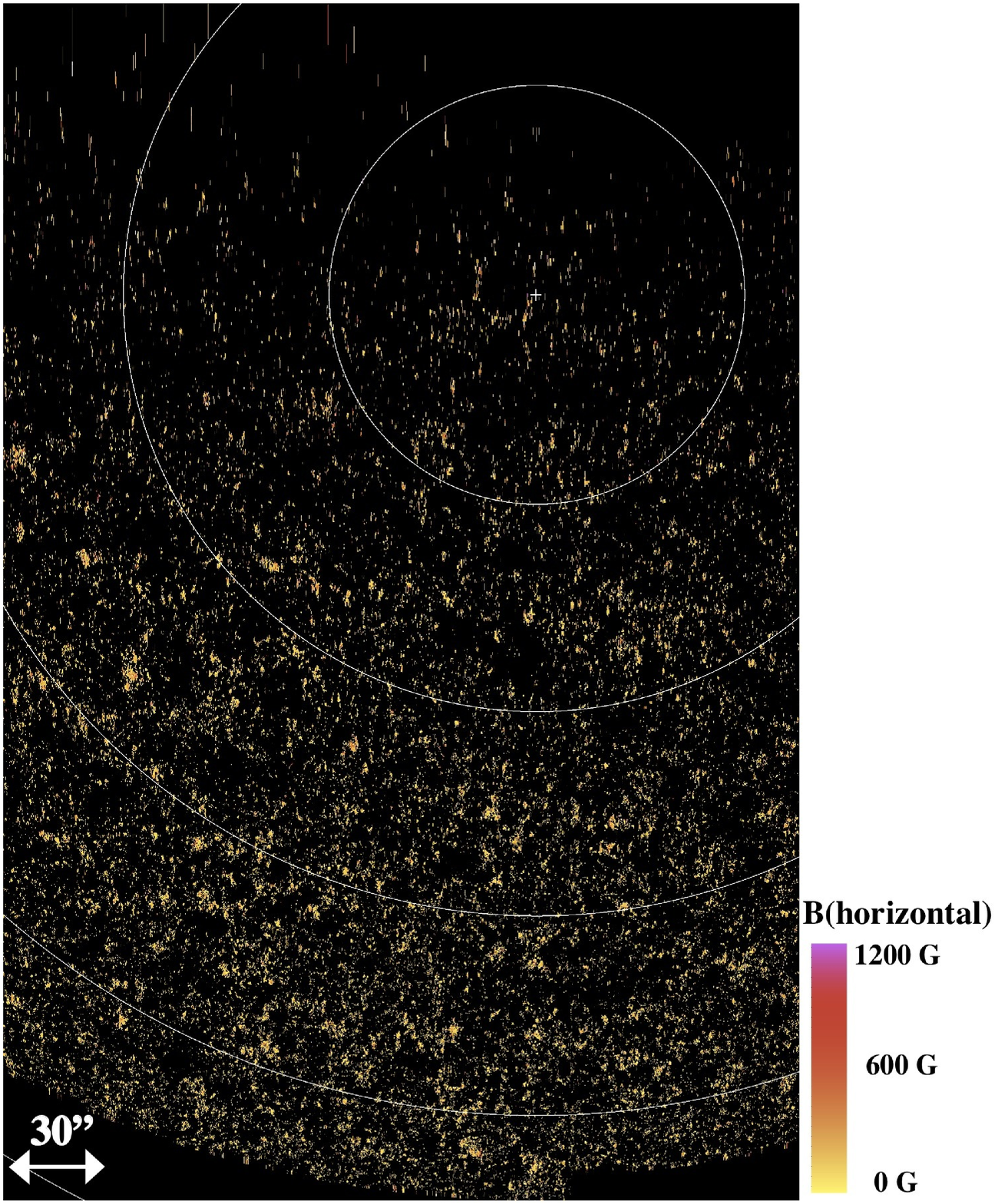}
 \end{center}
Fig. 4 ---{\it (b)} same as Figure 4 {\it (a)} but map of field strength of magnetic field vector classified as {\it horizontal}. 
 \label{npole3a}
\end{figure}

\begin{figure}
 \begin{center}
  \epsscale{1.00}
  \plotone{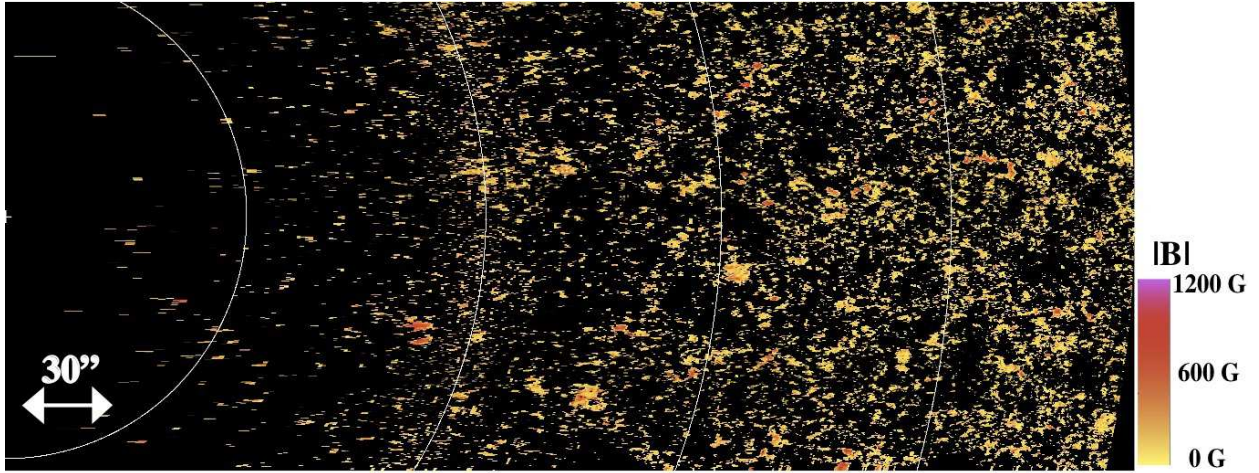}
 \end{center}
 \caption{
Magnetic landscape of the quiet Sun near the East limb; The map of the magnetic field strength on the sky plane is converted to the map seen from above the East limb. The observations are done at 21:00---23:16 UT on 2007 November 28. North is up. The pixel size is $0.16''$. The field of view is $100''$ (East-West) by $162.84''$ (North-South). The size of the field of view for East-West direction ($100''$) is expanded to  $397.6''$ as a result of correction for foreshortening. The circular arc is the latitudinal line per $5^\circ$ from the East limb. The magnetic field strengths were obtained only for pixels whose polarization signal exceeds a given threshold (see text). This figure can be directly compared with the polar map (Fig. \ref{npole2}). The scale size and the color table for these maps are the same. 
}
 \label{elimb2}
\end{figure}

\begin{figure}
 \begin{center}
  \epsscale{1.00}
  \plotone{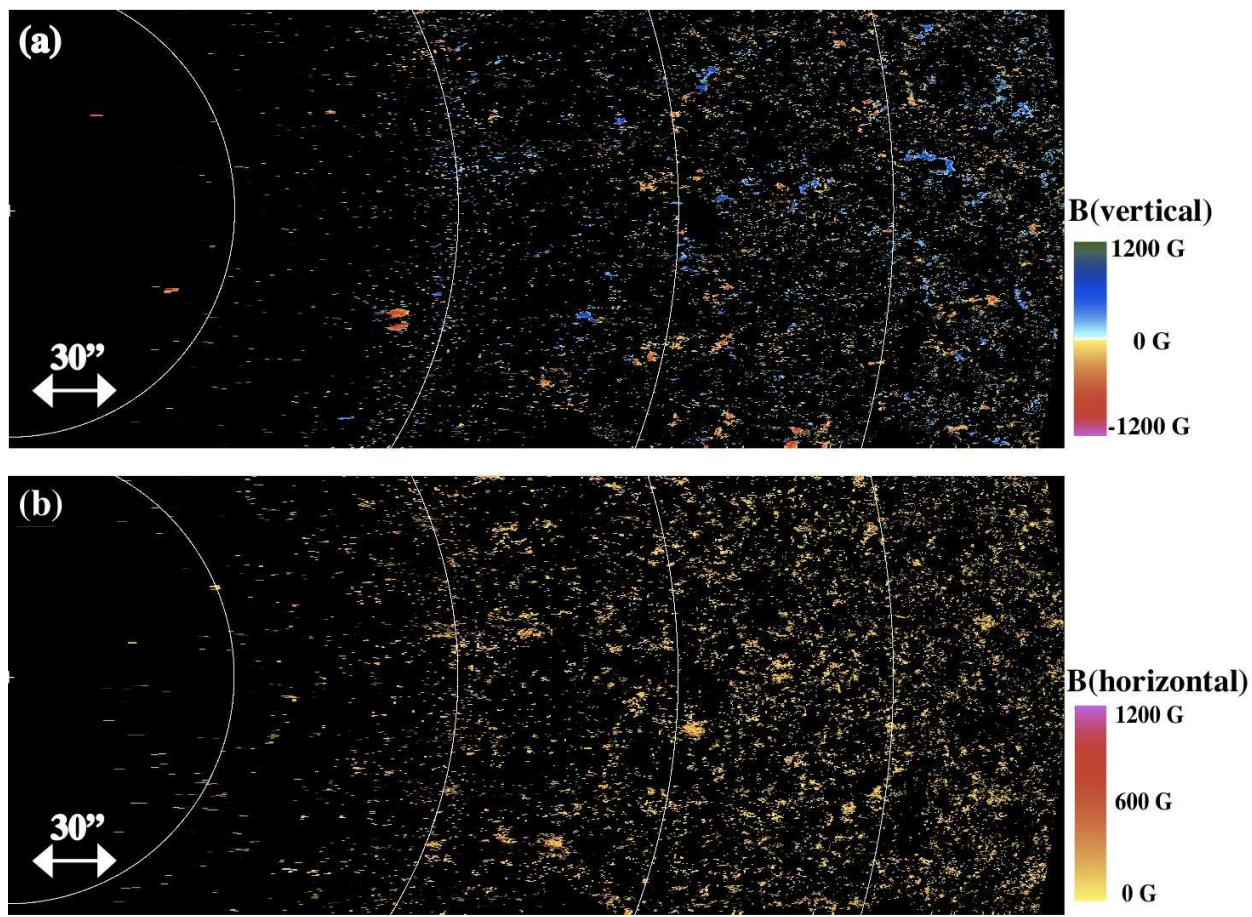}
 \end{center}
 \caption{
{\it (a)} Map of signed strength of the magnetic field vectors classified as {\it vertical} near the East limb. {\it (b)} Map of strength of the magnetic field vectors classified as {\it horizontal} near the East limb. The original data is the same as that of Fig. \ref{elimb2}. The pixel size is $0.16''$. The magnetic field strengths were obtained only for pixels whose polarization signal exceeds a given threshold (see text). These panels can be directly compared with the polar maps (Fig. \ref{npole3}). The scale size and the color table for these maps are the same.  
}
 \label{elimb3}
\end{figure}

\begin{figure}
 \begin{center}
  \epsscale{0.9}
  \plotone{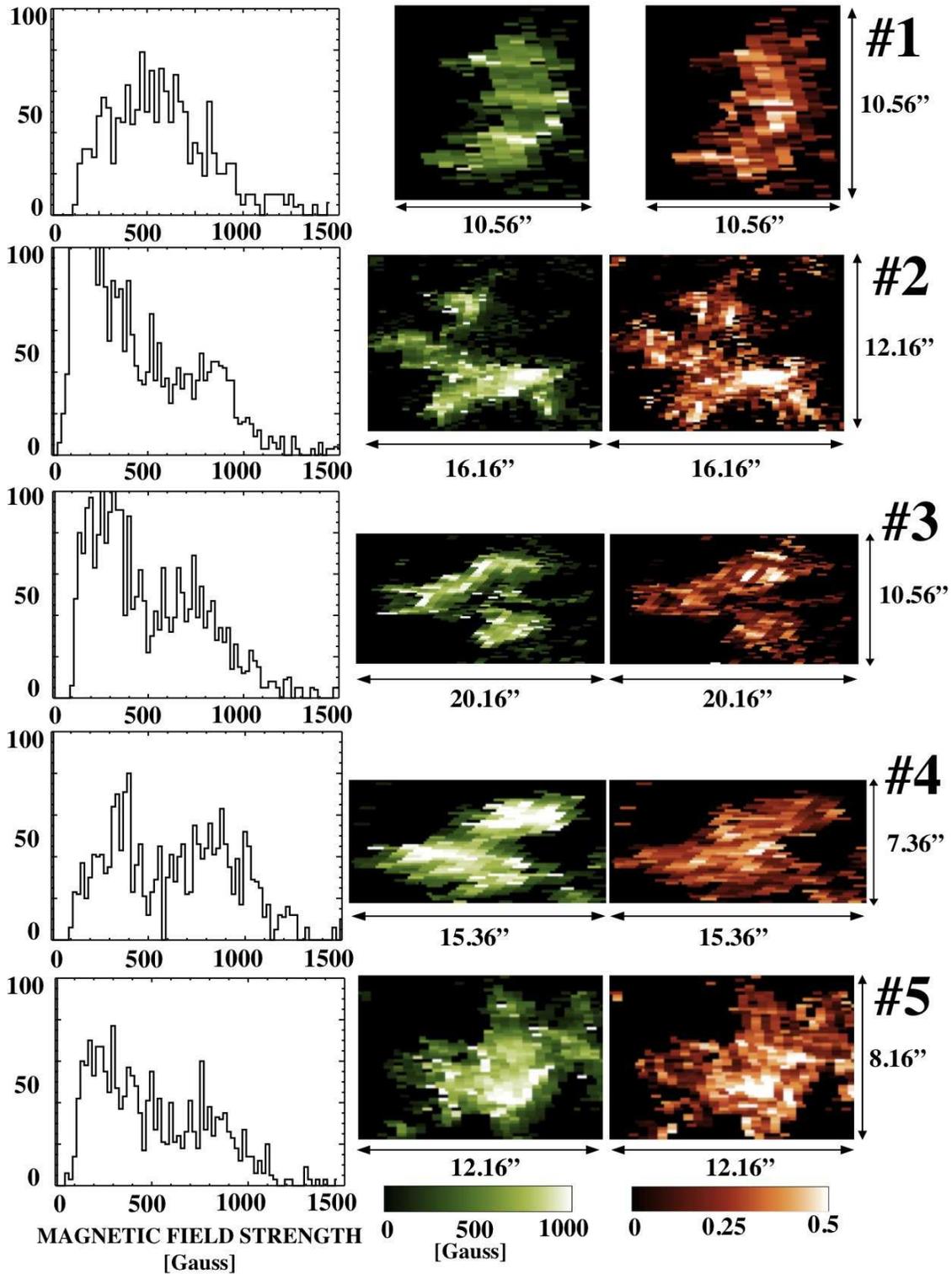}
 \end{center}
 \caption{
Data from the kG magnetic concentrations arbitrarily chosen in the North polar region (Fig. \ref{npole2}); the histograms of intrinsic magnetic fields strength (left), maps of intrinsic magnetic field strength (center) and the filing factor (right) seen from above the pole. The pixel size of the maps is 0.16$''$.
 }
 \label{npatch}
\end{figure}

\begin{figure}
 \begin{center}
  \epsscale{0.9}
  \plotone{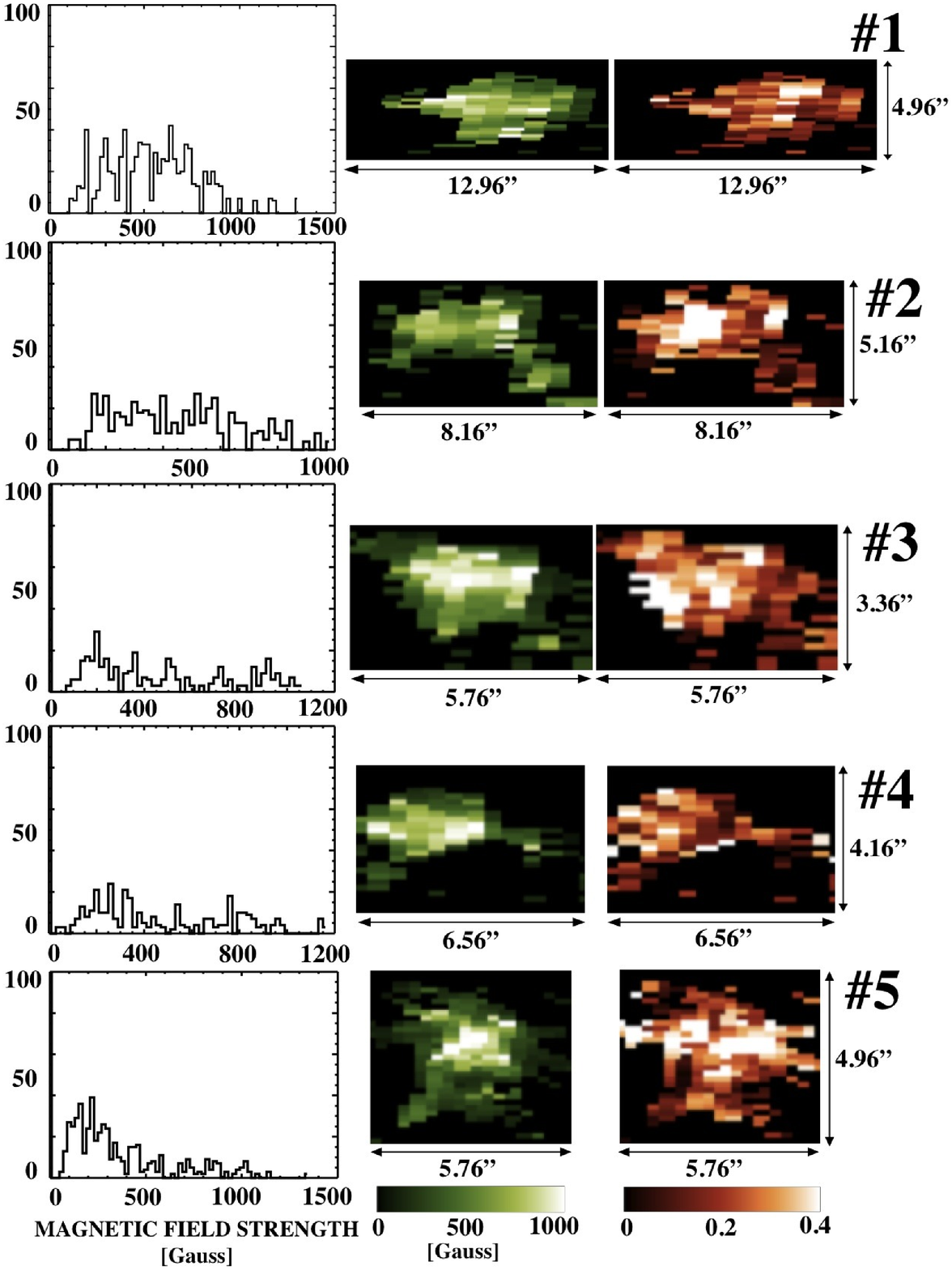}
 \end{center}
 \caption{ 
Data from the  kG magnetic concentrations arbitrarily chosen in the quiet Sun (Fig. \ref{elimb2}); the histograms of intrinsic magnetic fields strength (left), maps of intrinsic magnetic field strength (center) and the filing factor (right) seen from above the East limb. The quiet Sun has the kG-patches with both positive and negative magnetic polarities, and the absolute intrinsic magnetic field strengths are shown. The pixel size of the maps is 0.16$''$. 
 }
 \label{epatch}
\end{figure}

\section{OBSERVATIONS AND ANALYSIS}
{\it Hinode} observed the North polar region of the Sun on September 25, 2007, when the North pole was located $7^\circ$ inside the visible disk. The SP observed a field of view (FOV) of $320'' \times 163.84''$ in a very-deep-map mode (exposure time: 12.8 s). The SP records the Stokes spectral profiles ({\it I, Q, U} and {\it V}) of Fe I 630.15 nm and 630.25 nm with a wavelength sampling of 2.15 pm  ${\rm pixel}^{-1}$ and with a scan step of 0.16$''$. 

A least-squares fitting was applied to the Stokes spectra with the MILne-Eddington inversion of the pOlarized Spectra (MILOS) code \citep{david07}. The 10 free parameters are the three components describing the vector magnetic field (strength {\it B}, inclination angle $\gamma$, azimuth angle $\chi$ defined in Fig. \ref{incli1}), the line-of-sight velocity, two parameters describing the source function, the ratio of line-to-continuum absorption coefficient, the Doppler width, the damping parameter, and the stray-light factor $\alpha$. A nonzero stray-light factor $\alpha$ may be interpreted as a parameter that represents both the filling factor of a nonmagnetized atmosphere along the line of sight and the stray-light contamination. The stray-light profile is evaluated individually for each pixel as the average of the Stokes I profiles observed in a $1''$ wide box centered on the pixel \citep{david07b}. This arrangement also allows us to accurately estimate the stray-light profiles of rapidly changing continuum intensity toward the limb.  

To minimize the influence of noise leading to unreliable inversions, we analyze only pixels whose linear or circular polarization signal amplitudes exceed a given threshold above the noise level $\sigma$. The noise level $\sigma$ is determined in the continuum wavelength range of the Stokes {\it Q, U} and {\it V} profiles. The noise level $\sigma$ is given by  $\displaystyle \sigma = \sqrt{\sum_{i} (V_i-\overline{V})^2 /n}$, where $V_i$ is the intensity of the stokes V profile at continuum wavelength pixel $i$, $\overline{V}$ is the average Stokes V signal for the same wavelength range, and $n$ is the number of wavelength data points. $\sigma /I_c$ depends on the exposure time, where $I_c$ is the continuum intensity,  and is 0.0008 for the observation of the North polar region.  

The fitting was performed for pixels whose {\it Q, U} or {\it V} peak is larger than $5\sigma /I_c$. The derived maps still contain pixels with physically unacceptable values. If there is a strong spike on top of the noisy Stokes profiles (due to Poisson fluctuation and cosmic ray hits) exceeding the threshold, that pixel will pass the 5$\sigma$ criteria. This is the reason why we can not completely remove the bad pixels. Thus, we employ an additional filter to further remove the influence of the noise. We improved the statistics of  the Stokes profiles by summing across 2 pixels in wavelength, and then removed the pixels whose {\it Q, U} or {\it V} peak in the summed profiles is lower than $5\sigma/I_c$. All the good profiles that have passed the previous threshold should pass this criterion too: only the noisy profiles with high spike which have cheated the prior filter can be potentially removed in this stage. Note that we use the original Stokes profiles (not summed) for inversion. 

Furthermore, we remove pixels with the absolute value of the Doppler velocity higher than 10 km s$^{-1}$ {\it or} the filling factor less than 0.01. Analysis of several samples indicates that the Stokes profiles of such pixels are not fitted well. The velocity of the horizontal flow due to granular motion observed with SOT/SP is smaller than 9 km s$^{-1}$ \citep{Bellot09}, and the threshold of 10 km s$^{-1}$ employed here is above this value. If the stray-light contamination is negligible, the filling factor of the magnetic atmosphere will be given by $f=1-\alpha$ \citep{david07b}. \citet{tsuneta08} reported that the distribution of the filling factor $f$ has a broad peak at $f=0.15$ with FWHM range $0.05 < f < 0.35$ for another polar region, and the threshold given above for the filling factor is located far away from the typical range. Even if we erroneously remove valid pixels with this filter, the effect of the erroneous removal would be  negligible, since the total number of pixels removed with this particular process is only 1.7\% of the total number of pixels that passed the previous two criteria. A total of 21.0\% of the pixels pass all the thresholds.  

\section{NORTH POLAR REGION}
\label{polarregion}
Figure \ref{npole2} shows a map of the magnetic field strength as seen from just above the North pole. Such a polar projection is needed to correctly see the spatial extent and size distribution of magnetic patches in the polar region. Note that at a higher latitude the spatial resolution in one direction is compromised due to the projection effect. Large patches correspond to the kilo-Gauss magnetic fields with the fanning-out structure \citep{tsuneta08}. 

\subsection{Vertical and Horizontal Magnetic Fields}
The inclination (Zenith) angle of the magnetic field vector with respect to the local normal has a wide distribution with two peaks around the local vertical and the local horizontal directions \citep{david07a, ishi09}. Since we are interested in any difference between the polar region and the quiet Sun, and the resulting consequence in the coronal magnetic fields, we obtain the Zenith angle {\it i} of the magnetic field vector. Due to the $180^\circ$ ambiguity in the direction of the magnetic field vector projected onto the sky plane, there are two solutions ${\it i_{\rm 1}}$ and ${\it i_{\rm 2}}$ ($ 0 \leq {\it i_{\rm 1}} \leq \pi $, $ 0 \leq {\it i_{\rm 2}} \leq \pi$
; Fig. \ref{incli1}) given by
\begin{eqnarray}
 {\it i_{\rm 1}} &=& \arccos \{\cos \gamma \cos \theta  + 
  \sin \gamma \sin \theta \cos {\bf (\beta -\chi) } \}, \\
 {\it i_{\rm 2}}  &=& \arccos \{\cos \gamma \cos \theta  +
  \sin \gamma \sin \theta \cos {\bf (\pi-(\beta -\chi)) } \}, 
\end{eqnarray}
where $\theta$ is the angle between the local normal and the line-of-sight, and is given by $\theta = \cos^{-1} (\cos \delta \cos \varphi)$, where $\delta$ and $\varphi$  are heliographic latitude and longitude for a pixel, respectively, $\gamma$ is the inclination of the magnetic field vector with respect to the line of sight ($ 0 \leq \gamma \leq \pi $), and $\chi$ is the azimuth angle ($ 0 \leq \chi \leq \pi $). The $\beta$ angle ($ 0 \leq \beta \leq \pi $) is defined in Fig. \ref{incli1}, and is given by $\beta = \cos^{-1} (\cos \delta \sin \varphi / \sin \theta)$.

We now introduce one assumption that the magnetic field vector is either {\it vertical} or {\it horizontal} to the local surface (or undetermined). The Zenith angle is defined to be from $0^\circ$ to $40^\circ$ and from $140^\circ$ to $180^\circ$ for the {\it vertical} magnetic field. The Zenith angle of the {\it horizontal} field is defined to be between $70^\circ$ and $110^\circ$, following \citet{ishi09}. Magnetic field vectors with an inclination angle  between 40$^\circ$ and 70$^\circ$ are not used in the subsequent analysis. According to this definition, every pixel has magnetic field vector classified as either {\it vertical} or {\it horizontal} (or undetermined). Figure \ref{incli2} shows the scatter plots of the two solutions ${\it i_{\rm 1}}$ and ${\it i_{\rm 2}}$ for the Zenith angle. If both the solutions are vertical, the one closer to the local normal is taken. In case one solution is vertical and the other horizontal, it is not possible to distinguish one from each other. Pixels with those solutions are not used. If one of the solutions is either vertical or horizontal, and the other solution is neither vertical nor horizontal, we will choose the solution of either vertical or horizontal. In addition, the kG-patches with quasi-symmetric fanning-out structure allows us to manually correct the Zenith angle by visually inspecting the thus-determined Zenith angle map.

There is concentration of pixels (about 3\% of the pixels) along the white line of ${\it i_{\rm 1}} \sim {\it i_{\rm 2}}$ in Figure \ref{incli2}. These pixels along the line have only valid Stokes V signal without Stokes Q and U signals. The inversion code correctly produces either $\gamma \sim 0$ or $\gamma \sim \pi$ for these pixels, so that ${\it i_{\rm 1}} \sim {\it i_{\rm 2}}$ from Equations (1) and (2). There is a gap in population around the line ${\it i_{\rm 1}} + {\it i_{\rm 2}} \sim \pi$. If the inclination angle with respect to the line of sight is  $\gamma \sim \pi/2$, i.e. the magnetic field vector is on the sky plane, it turns out that ${\it i_{\rm 1}} + {\it i_{\rm 2}} \sim \pi$ from Equations (1) and (2). There may be two reasons for the gap: (1) The noise in Stokes V makes the derived magnetic vector impossible to reach an inclination of $90^\circ$ (ie. the sky plane). (2) Stokes V is much more sensitive than Stokes Q and V for weak magnetic field strength, so that the magnetic field vectors tend to deviate the sky plane. 

It is possible to check the consistency of the $180^\circ$ ambiguity resolution done here with the information on the sign of the Stokes V signal \citep{tsuneta08}. We compare the polarity of the magnetic field vectors classified as {\it vertical} shown in Figure \ref{npole3} with the Stokes V map such as the one shown in Figure 2 of \citet{tsuneta08}, and confirm that 96\% of the pixels shown in Fig. \ref{npole3} has the correct sign indicated by the polarity of the Stokes V signal. In particular, we confirm that all the apparent kG-patches have the correct sign. This indicates that the assumptions and the rules employed to classify the {\it vertical} and {\it horizontal} magnetic field vectors are valid.   

\subsection{North Polar Region}
Two maps of the strengths of the magnetic field vectors classified as {\it vertical} and {\it horizontal} for the North polar region are shown in Figure \ref{npole3}. Comparison of Figure \ref {npole2} with two panels of Figure \ref{npole3} shows that the vertical and horizontal magnetic fields are well decoupled: For instance, we do not see any inclined fanning-out structure around the magnetic concentrations of the vertical magnetic fields in the map of the horizontal magnetic fields due to the separation band prepared in between the vertical and horizontal zones in the inclination of the magnetic field vectors. 

As for the vertical magnetic fields, almost all the large patches have the same polarity, while both polarities exist for the smaller patches. There appears to be more mixed polarities in lower latitude, while the region in higher latitude is dominated by negative kG patches. The horizontal magnetic fields appear to be much more uniform in size and in spatial distribution as compared with the vertical magnetic fields.   

\section{QUIET SUN AT THE EAST LIMB}
The quiet Sun at the East limb was observed on November 28, 2007 using the SOT/SP in the very-deep-map mode. The exposure time is 12.8 s, which is the same as that of the polar observation. We confirm with that there is no coronal activity (enhanced brightness) in the region with the data from the X-ray telescope aboard {\it Hinode} . The $\sigma /I_c$ for the quiet Sun at the East limb is 0.0008. The signal-to-noise ratio is the same as that for the North polar region, simply because the exposure time and the observing mode for the two observations are the same. The size of the FOV was $100'' ({\rm EW}) \times 162.84'' ({\rm NS})$. As we did for the North polar region, a least-squares fitting is applied to the Stokes spectra with the MILOS code. Figure \ref{elimb2} shows the map of the magnetic field strength as seen from just above the East limb to allow us to compare the spatial extent and size distribution of magnetic patches in the quiet Sun and the polar region. In the region closer to the limb, the spatial resolution in one direction is reduced due to the projection effect just like in the North pole region. Similarly, there are two solutions of the Zenith angle for the magnetic field vectors. We have followed the same assumptions that were employed for the North pole. Two maps of the strengths of the magnetic field vectors classified as {\it vertical} and {\it horizontal} are shown in Figure \ref{elimb3}. 

The spatial distribution of the vertical magnetic fields at the East limb appears to be different from those in the North pole. The map of the quiet Sun shows the mixture of magnetic patches with both polarities. The magnetic concentrations at the East limb are apparently smaller in size than those at the North pole. The number of the patches is also smaller in this particular quiet Sun region. Contrastingly the horizontal magnetic fields of the two regions appear to be strikingly similar.  

\section{COMPARISON BETWEEN THE POLAR REGION AND THE QUIET REGION OF THE SUN} 

\subsection{Kilo-Gauss Magnetic Patches}
Figures \ref{npatch} and \ref{epatch} show arbitrarily chosen kG-patches in the North polar region and in the quiet Sun at the East limb. These patches are located between $0^\circ$ to $23^\circ$ from the pole and the East limb.  We notice complex internal structures of the kG patches \citep{Oku04,Oku05}. The histograms of the magnetic field strengths for the two regions are apparently different. The magnetic concentrations in the polar region have an enhancement at around 800 G, while in the quiet Sun such deviations are not apparent. The statistics is poor in the quiet Sun, and this could be the reason why no deviation is seen. The intrinsic magnetic field strengths are at their maximum value around the center of the kG-patches, and decrease towards the edge. Likewise, the filling factors essentially reach a maximum around the center of the kG-patches, and decrease toward the edges. The true areas, the maximum magnetic field strengths, and the average field strengths of each kG-concentration shown in Figures \ref{npatch} and \ref{epatch} are shown in Table. \ref{patch_property} (from \#1 to \#5). 

\begin{deluxetable}{cccc}
\tablecolumns{4}
\tablewidth{0pc}
\tabletypesize{\scriptsize} 
\tablecaption{Properties of kG-patches}
\tablehead{
\multicolumn{4}{c}{North polar region} \\
\cline{1-4} \\
\colhead{} & \colhead{Area} & \colhead{Maximum field strength} &
\colhead{Average field strength} \\
\colhead{number} & \colhead{[$\rm{cm}^{2}$]} & \colhead{[Gauss]} &
\colhead{[Gauss]}
}
\startdata
\#1 & 2.1$\times 10^{17}$ & 1166 & 460 \\
\#2 & 4.0$\times 10^{17}$ & 2217 & 444 \\
\#3 & 3.6$\times 10^{17}$ & 1657 & 504 \\
\#4 & 2.7$\times 10^{17}$ & 1678 & 638 \\
\#5 & 2.6$\times 10^{17}$ & 1461 & 507 \\
\#6 & 1.6$\times 10^{17}$ & 2052 & 529 \\
\#7 & 1.6$\times 10^{17}$ & 1363 & 430 \\
\#8 & 1.3$\times 10^{17}$ & 1942 & 524 \\
\#9 & 1.5$\times 10^{17}$ & 1370 & 713 \\
\#10 & 2.1$\times 10^{17}$ & 1764 & 414 \\
\hline
Average & 2.3$\times 10^{17}$ & 1648 & 516 \\
\cutinhead{East limb}
\#1 &11.3$\times 10^{16}$ & 1290 & 549 \\
\#2 & 7.2$\times 10^{16}$ &  982 & 452 \\
\#3 & 5.1$\times 10^{16}$ & 1063 & 481 \\
\#4 & 4.6$\times 10^{16}$ & 1167 & 477 \\
\#5 & 7.5$\times 10^{16}$ & 1336 & 373 \\
\#6 & 7.2$\times 10^{16}$ & 1242 & 522 \\
\#7 & 5.7$\times 10^{16}$ &  805 & 351 \\
\#8 & 6.1$\times 10^{16}$ & 1139 & 401 \\
\#9 & 4.2$\times 10^{16}$ & 1049 & 355 \\
\#10 & 9.7$\times 10^{16}$ & 1917 & 440 \\
\hline
Average & 6.9$\times 10^{16}$ & 1179 & 440 \\
\enddata
\label{patch_property}
\end{deluxetable}

To investigate the properties of typical large kG-patches, we chose additional 5 large patches at latitudes between $75^\circ$ to $90^\circ$ in the North polar region (Fig. \ref{npole3}). These are added to Table. \ref{patch_property} (from \#6 to \#10). (In total, we have 10 data sets.) The average maximum intrinsic magnetic field strength is 1600 G, and the average intrinsic magnetic field strength is 500 G. The average area is about $2.3\times 10^{17} {\rm cm^{2}}$. The total magnetic flux of a kG-patch is estimated with $\sum B_j \cos {\it i}_j f_j s_j$, where $B_j$, ${\it i}_j$ and $f_j$ are the intrinsic magnetic field strength, the inclination angle obtained with Equations (1) and (2), and the filling factor of the {\it j}-th SOT pixel inside the kG-patch, respectively, and $s_j$ is the pixel size. Average magnetic flux of the ten large patches in the polar region is $2.0 \times 10^{19}$ Mx per patch. These fluxes estimated here and below are the minimum values in the sense that the sizes of the patches are determined with the $5\sigma$ threshold.  

We also include additional 5 large patches located in the belt between $67^\circ$ to $90^\circ$ (Fig. \ref{elimb3}) in the quiet Sun at the East limb (Table. \ref{patch_property}; from \#6 to \#10). The average maximum intrinsic magnetic field strength is 1200 G, and the average intrinsic magnetic field strength is 400 G. The average area is $6.9 \times 10^{16} {\rm cm^{2}}$. Average magnetic flux of the ten large patches in the quiet Sun is $4.2 \times 10^{18}$  Mx per patch. The intrinsic magnetic field strengths of the polar magnetic patches are larger than those of the quiet Sun. The average area and the total magnetic flux have considerable difference between the two regions: kG magnetic concentrations in the polar region have a factor of 3.3 larger average area, and have a factor of 4.8 larger magnetic flux than those in the quiet Sun. 

\subsection{Histogram of Magnetic Field Strength}

\begin{figure}
 \begin{center}
  \plotone{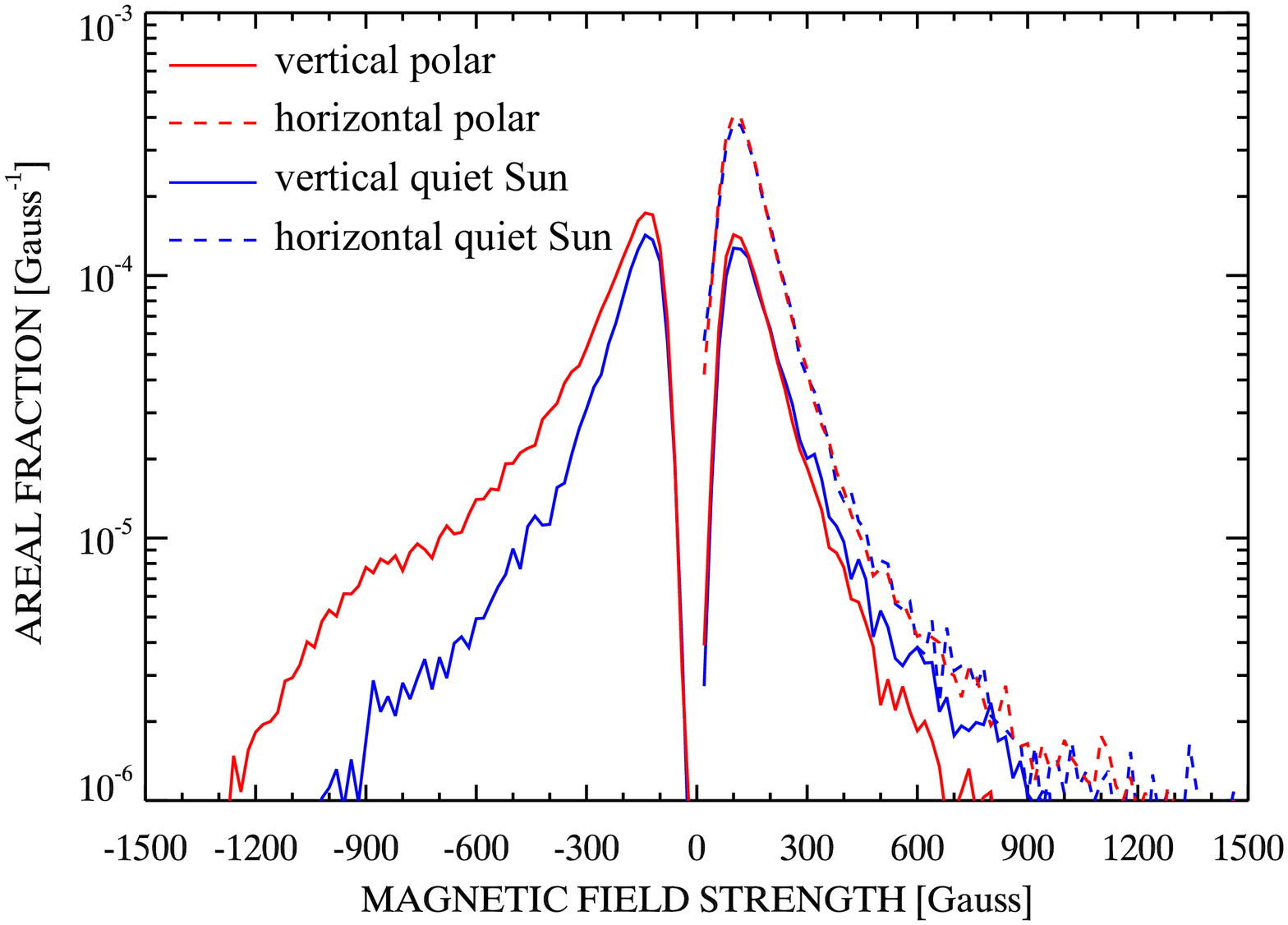}
 \end{center}
 \caption{
Areal fraction of the intrinsic magnetic field strength. The red and blue lines represent the North polar region and the quiet Sun at the East limb, respectively. The solid and dashed lines represent the magnetic fields classified as {\it vertical} and {\it horizontal} in this work (see Sect. 3.1 for details), respectively . Vertical magnetic field has a sign that indicates either plus or minus polarities, while horizontal magnetic field does not have sign. Vertical axis is the number of pixels divided by total number of pixels (including the pixels for which inversion is not performed) in the respective field of view.
}
 \label{PDF}
\end{figure}

\begin{figure}
 \begin{center}
  \plotone{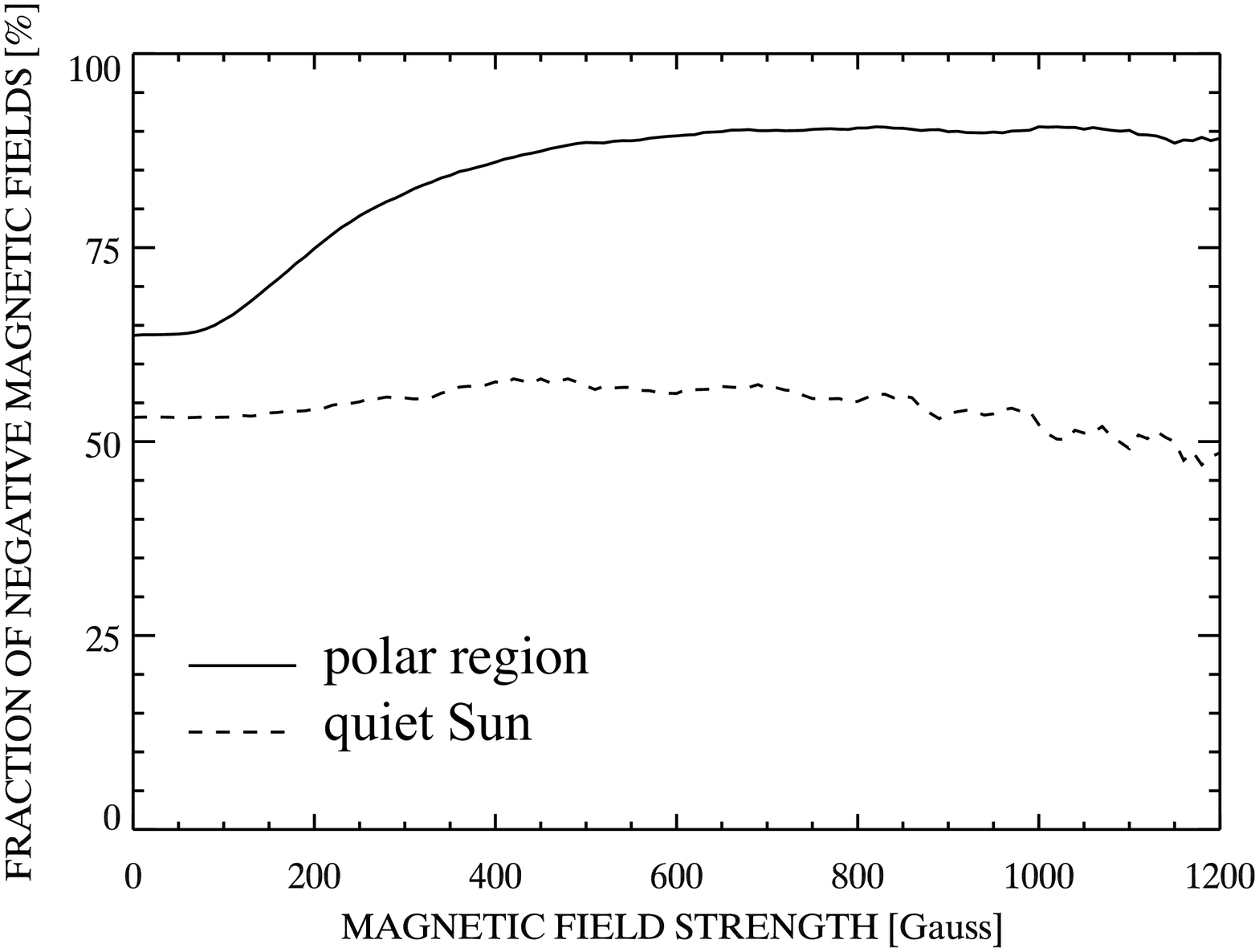}
 \end{center}
 \caption{
Fractional number of pixels that have negative vertical magnetic field with the intrinsic strength stronger than the value in the horizontal axis; North polar region (bold line) and the quiet Sun at East limb (broken line). 
}
 \label{polar_balance}
\end{figure}

We observed the quiet Sun located at the extreme East limb to provide a reference data set for the polar region. Fig. \ref{PDF} shows the areal fraction (histogram) of the intrinsic magnetic field strength. The histograms are normalized to the total number of pixels in each SOT/SP map including those pixels for which Milne-Eddington inversion is not carried out, so that we can directly compare the two distributions. There is a large difference in the distribution of the magnetic fields classified as {\it vertical}: The distribution for the quiet Sun at the East limb is symmetric around zero, clearly showing balanced polarity in the magnetic flux, while the distribution for the North polar region is highly asymmetric, showing a predominance of a single polarity. The distributions for the positive polarity are similar between the quiet Sun and the polar region, while in the negative polarity regime, the amount of the flux is higher in the polar region than in the quiet Sun. For instance, fractional number of pixels with intrinsic field strength of $-$1 kG in the polar region is about 5 times larger than that in the quiet Sun. 

We also point out that the distributions of the magnetic fields classified as {\it horizontal} \citep[][references therein]{ishi09} in both the regions are strikingly similar in spite of the considerable difference in the amount of the {\it vertical} magnetic fields between the quiet Sun and the polar region. 

Figure \ref{polar_balance} shows fraction of the negative {\it vertical} magnetic fields with the intrinsic field strength stronger than the value in the horizontal axis. The fraction at 0 G indicates that the area covered by the negative polarity is 63.7 \% (polar region) and 53.1 \% (quiet Sun). The polar region is more dominated by negative polarity magnetic field with increasing field strength, while in the quiet Sun, magnetic fields of both polarities are equally distributed independent of field strength. In the polar region, the fraction of the negative magnetic flux reaches about 90 \% at 1 kG. Clearly bipolar fields are dominant in the quiet Sun, while the unipolar fields are predominant above 500 G in the polar region.  

The total magnetic flux for the magnetic field vectors classified as {\it vertical} in the entire field of view is defined to be $\sum B_j \cos i_j f_j s_j$, where $B_j$, $i_j$ and $f_j$ are the intrinsic magnetic field strength, the inclination angle obtained with Equations (1) and (2), and the filling factor of the {\it j}-th SOT pixel inside the SOT/SP field of view, respectively, and $s_j$ is the pixel size. The true size of the field of view is $8.4 \times 10^{20}$ cm$^{2}$ (polar region) and $3.5 \times 10^{20}$ cm$^{2}$ (East limb). The total magnetic flux in the SOT/SP field of view is $1.7 \times 10^{21}$ Mx for the polar region, and $4.0 \times 10^{20}$ Mx for the quiet Sun. (Note the factor of 2.4 difference in the observed area between the polar region and the quiet Sun.) 

The total magnetic flux for the magnetic field vectors classified as {\it horizontal} is more difficult to estimate. To make a crude estimate, we define it to be $\sum B_j \sin i_j L_j H_j$ \citep{ishi10}, where $B_j$, and $i_j$ are the intrinsic magnetic field strength and the inclination angle obtained with Equations (1) and (2) respectively \citep[e.g., Figure 9 of][]{ishi10}. $L_j$ is the linear size of the pixel size, and is commonly 0.16$''$. $H_j$ is the diameter of the horizontal flux tubes, and is commonly assumed to be 190 km, following \cite{ishi10}. Here the filling factor $f_j$ is replaced with the representative diameter $H_j$ of the horizontal flux tubes. The diameter is smaller than the thickness of the line forming layer, contributing to the apparent filling factor $f_j <1$ for horizontal magnetic fields. The total magnetic flux thus estimated is $9.9 \times 10^{21}$ Mx for the polar region, and $4.0 \times 10^{21}$ Mx for the quiet Sun. The ratio in magnetic flux ({\it horizontal/vertical}) is 10 (quiet Sun) and 5.8 (polar coronal hole), and the magnetic flux classified as horizontal is larger than that of the magnetic flux classified as vertical by a factor 5.8--10 \citep[e.g.,][]{Lites08}. 

The present result also shows that the estimation of the radial magnetic flux in the polar regions, if measured only with longitudinal magnetographs, may be contaminated by the massive presence of the horizontal magnetic fields.
 
\section{Coronal Magnetic Field Structure}
\begin{figure*}
 \begin{center}
  \epsscale{1.}
  \plotone{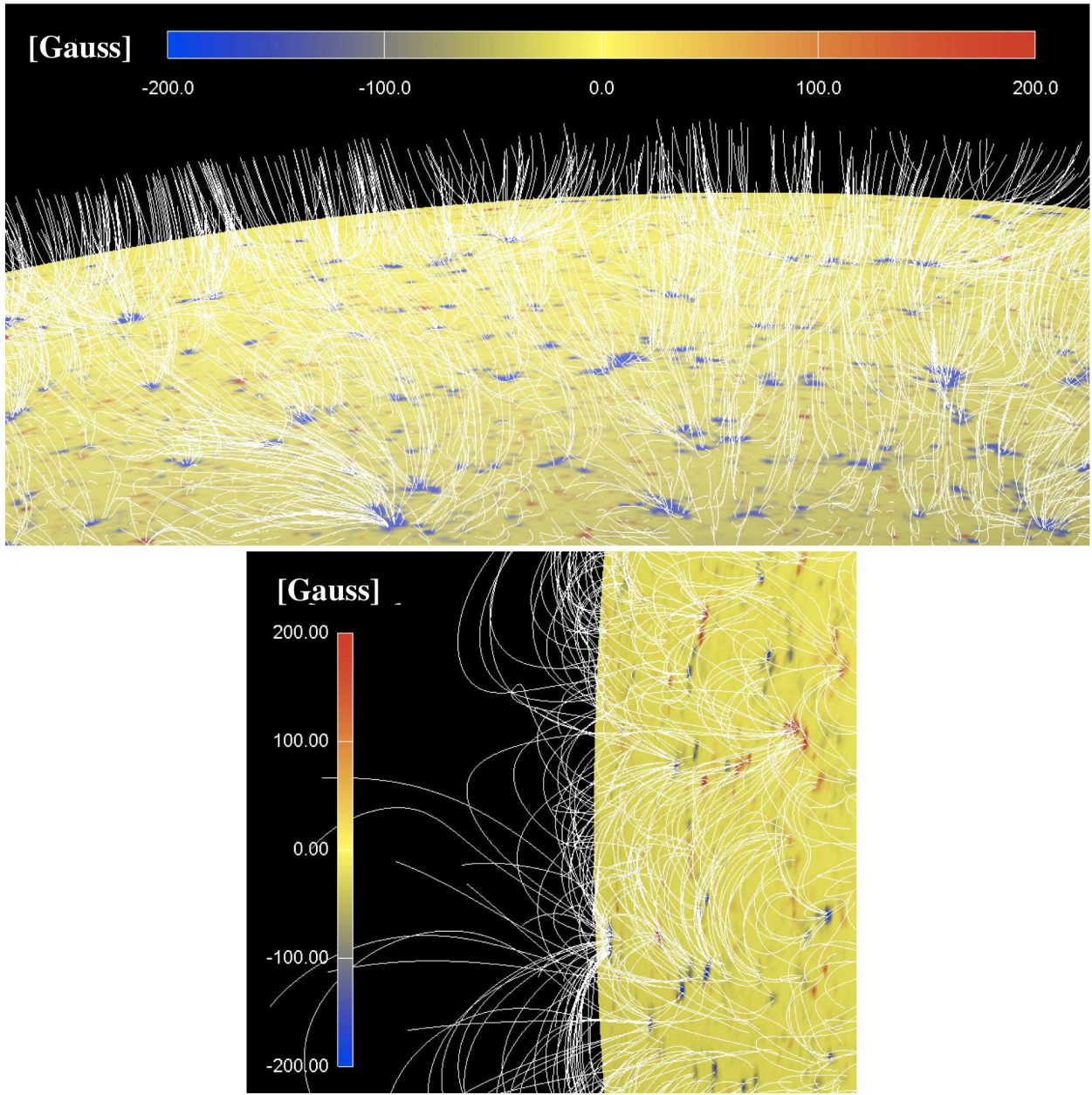}
 \end{center} 
 \caption{
Coronal magnetic field structure with the data shown in Figs. \ref{elimb3} and \ref{npole3} for the quiet Sun at the East limb (lower panel), and the North polar coronal hole (upper panel) respectively. The color of the magnetic patches indicates intrinsic field strength of the magnetic field vectors classified as vertical. The red patches are positive magnetic concentrations, and the blue ones are  negative.
}
 \label{pfss}
\end{figure*}

\begin{figure}
 \begin{center}
  \plotone{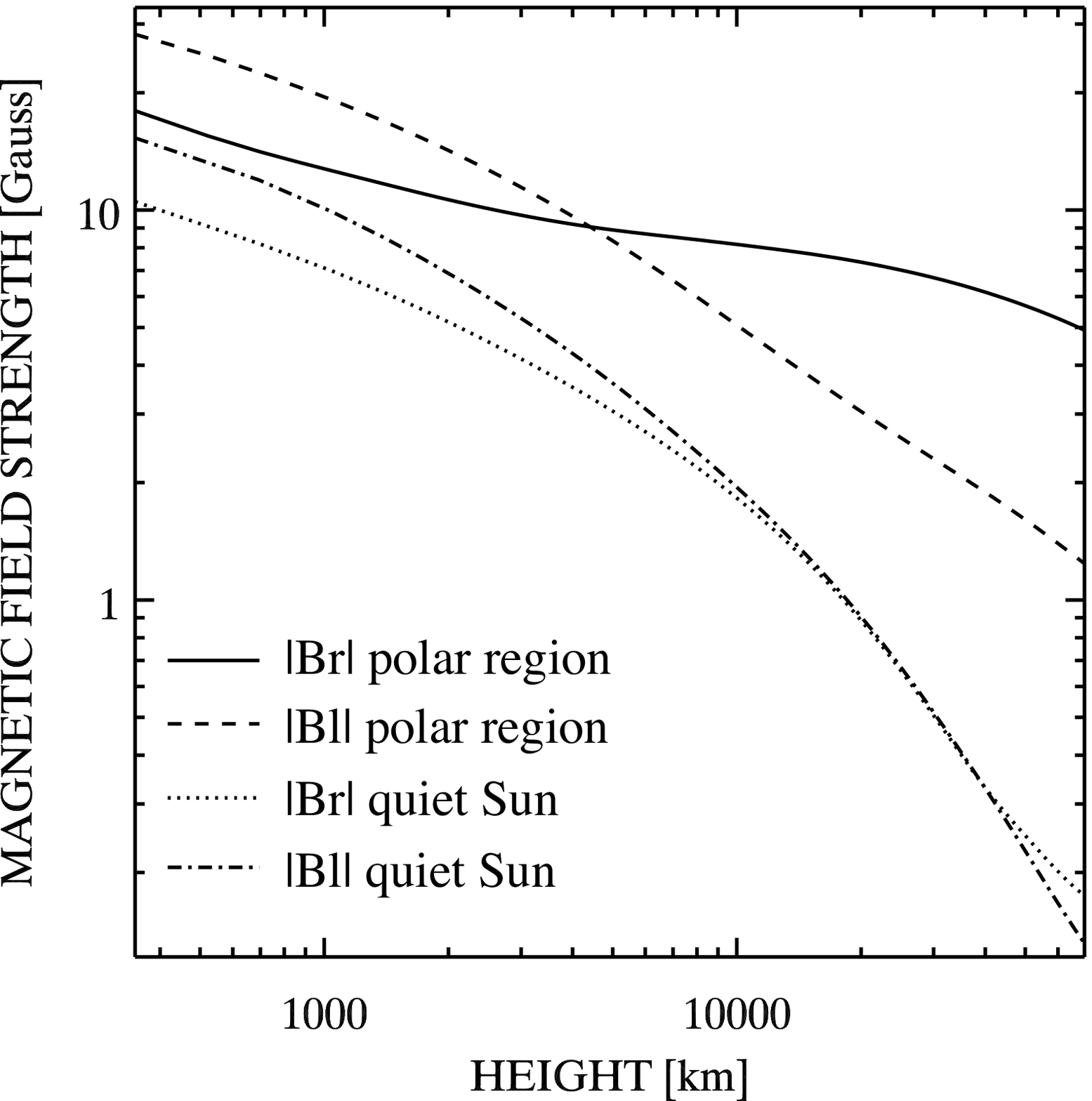}
 \end{center}
 \caption{
Average magnetic field strength as a function of height for the North polar region and the quiet Sun at the East limb. The dashed and dash-dotted lines represent the lateral component $|Bl|$ of the average magnetic field, and the solid and dotted lines indicate the radial component $|Br|$ at each height. 
}
 \label{BrBh}
\end{figure}

A three dimensional coronal magnetic field is reconstructed using a high resolution potential field source surface (PFSS) model recently developed by \citet{Shiota09}. The {\it Hinode} vertical magnetic field data are embedded in the high resolution {\it MDI} synoptic maps\footnote{http://soi.stanford.edu/magnetic/index6.html} of the Carrington Rotation 2061 (for the polar region) and 2063.35 (for the quiet Sun at the East limb). The size of the {\it MDI} maps is (3600, 1080) pixels. The  synoptic map at the Carrington Rotation 2063.35 is the combined map of the Carrington Rotations 2063 and 2064 such that the embedded {\it Hinode} region is located in the center of the synoptic map. We assume that the filling factor of the {\it Hinode} data is equal to 1, and that the magnetic field vectors classified to vertical are vertical relative to the local photospheric normal. The synoptic maps are expanded with spherical harmonics of extremely high degree ($L_{\rm max}=2048$) whose equatorial resolution is $\sim$ 1072 km.

Figure \ref{pfss} shows the snapshots of the inferred three dimensional coronal magnetic fields reconstructed with the PFSS model where the source surface is assumed to be located at 2.5 R$_\odot$. We can clearly see that the majority of magnetic field lines from the kG-patches in the polar coronal hole are open with canopy structures just above the photosphere. There are some closed magnetic field lines in the polar coronal hole as well. On the contrary, in the quiet Sun, almost all the field lines are closed. (We confirm that some field lines going outside the box are connected to other parts of the Sun.) The bipolar nature of the quiet Sun, and the unipolar nature of the polar coronal hole is one of the significant differences between the two regions. 


Alfv\'en waves are believed to be responsible for the acceleration of the fast solar wind \citep{Suzuki06}. Whether the Alfv\'en waves generated in the photosphere are reflected back \citep{fuji} at the photosphere/corona boundary may depend on the degree of the expansion of the flux tubes as a function of height. Open field lines from the kG-patches in the polar region may expand at very low altitude, so that there will be less discontinuity in the Alfv\'en velocities. Magnetic flux tubes with fanning-out structures in the very low atmosphere may serve as a chimney for the Alfv\'en waves to propagate to the corona \citep{tsuneta08}. 

Figure \ref{BrBh} shows the average magnetic field strength as a function of height in the North polar region and the quiet Sun at the
East limb. $|Br|$ and $|Bl|$ represent the averaged radial and lateral components of the coronal magnetic field obtained by the high resolution PFSS model, respectively, where $|B| = \sqrt{Br^2 + Bl^2}$. The radial magnetic field strength decreases with altitude significantly slower in the polar region than in the quiet Sun, and magnetic field strength above the polar region is much higher than that in the quiet Sun.
 
\section{DISCUSSION AND CONCLUSION}
In this paper, we have made a detailed comparison of the photospheric magnetic properties between the polar region and the quiet region. We found that the magnetic properties of the polar region are significantly different from those of the quiet Sun. Positive and negative magnetic fields are well balanced in the quiet Sun, while in the North polar region, negative-polarity magnetic fields dominate the other polarity. The excess negative magnetic field makes both total unsigned magnetic flux and net (signed) magnetic flux larger than those of the quiet Sun.  

\subsection{Kilo-Gauss Magnetic Patches}
All the large magnetic concentrations in the polar region essentially have the same magnetic polarity, while the smaller patches have the same and the opposite magnetic polarities. \citet{Shimojo09} reported bipolar emergence of magnetic fields in the polar region, and this may contribute to the conspicuous minority polarity patches. The average area and the average magnetic flux of the kG patches appear to be considerably different between the two regions: the sample data set tabulated in Table 1 shows that kG magnetic concentrations in the polar region have a factor of 3.3 larger average area, and have a factor of 4.8 larger total magnetic flux than those in the quiet Sun. The number of the kG-patches in the polar region appears to be larger than that of the quiet Sun. 

The origin and/or evolution of the kG-patches would be different between the polar region and the quiet Sun. The different properties of the kG-patches in the polar regions and the quiet Sun could be due to the different environment in which they evolve. There may be a higher chance in the quiet Sun that the positive and negative patches collide, reconnect, and lose magnetic energy or submerge as a result. As such, the environment in the quiet Sun may not allow the elemental magnetic concentrations to grow. 

\subsection{Horizontal magnetic fields}
From Fig. \ref{PDF}, 98\% of the horizontal magnetic fields has the intrinsic field strengths smaller than 700 G both in the quiet Sun at the limb and the polar region. Here the horizontal magnetic fields refer to the magnetic field vectors classified as horizontal. \cite{ishi09} reported that 93\% of horizontal magnetic fields have intrinsic field strengths smaller than 700 G, and 98\% smaller than 1 kG in the quiet Sun and a plage region both located near the center of the solar disk. Magnetic field strength of 700 G corresponds to the typical equipartition field strength at the level of granules, where the density is $10^{-6}$ g cm$^{-3}$ at the depth of 500 km and the velocity is 2 km s$^{-2}$. If we take the mean density $2.7 \times 10^{-7}$ g cm$^{-3}$ at the base of the photosphere, the equipartition field strength is 350 G. Even in this case, 92\% of the horizontal magnetic fields has the intrinsic field strengths smaller than 350 G both in the quiet Sun at the limb and the polar region. Although the vertical magnetic flux in the polar region is larger than that of the quiet Sun, the horizontal magnetic flux in the polar region is remarkably similar to that of the quiet region. The small difference between the two results (98\% in the present paper and 93\% in \cite{ishi09} at 700 G) could be due to the following two factors: (1) A higher atmosphere is observed in the limb observations, so that the measured magnetic field strength may be smaller in the limb observations. (2) The horizontal magnetic fields are observed through Stokes Q and U in the disk center observations, while Stokes V is involved in the limb observations. Thus, the limb observations may have higher sensitivity to the horizontal magnetic fields. 

\cite{ishi09} proposed that a local dynamo process be uniformly operating in the photosphere or below based on the observations that the probability distribution function of the transient horizontal magnetic fields in the quiet Sun is the same as that in a plage region. If we assume that the magnetic properties of the quiet Sun in the disk center analyzed by \cite{ishi09} are the same as those of the quiet Sun at the East limb analyzed in this paper, we would conclude that the distribution function of the horizontal magnetic fields and the amount of the magnetic flux are exactly the same in the quiet Sun, a plage region and in the polar region. This provides firmer evidence that a local dynamo process driven by the granular motion is responsible for the generation of the horizontal magnetic fields.    

\subsection{Polar coronal holes}
\citet{Kano08b} obtained the temperatures and the densities of the polar coronal hole and the quiet Sun with the X-ray telescope (XRT) \citep{Golub07,Kano08a} aboard {\it Hinode}. These are the most accurate measurements carried out using  the broadband filters by correcting for the effect of scattered X-rays with the lunar occultation. The temperatures and electron densities then derived are 1.0 MK and 1.0$-$1.5 $ \times 10^8 {\rm cm^{-3}}$ for the polar coronal hole, and 1.5$-$2.0 MK and 2.0$-$3.0 $\times 10^8 {\rm cm^{-3}}$ for the quiet Sun. The plasma pressures are 1.4$-$2.1 $\times 10^{-2} {\rm dyne \cdot cm^{-2}}$ and 4.1$-$8.3 $\times 10^{-2} {\rm dyne \cdot cm^{-2}}$ for the coronal hole and the quiet Sun, respectively. Thus, plasma pressure in the coronal hole is a factor of 3$-$6 smaller than that of the quiet Sun.  

The coronal hole in the upper corona inside the Alfv\'enic sphere should have collapsed due to the lateral pressure from the surrounding quiet Sun corona, if the magnetic field strength in the polar coronal hole is much smaller than that of the quiet corona. The polar coronal hole is stable for a long time, and the coronal hole and the surrounding corona of the quiet Sun must be in pressure equilibrium. In this paper, we showed that the magnetic flux of the polar region in the photosphere is larger than that in the quiet Sun (Fig. \ref{PDF}). Indeed, magnetic pressure estimated from Fig. \ref{BrBh} is 1 dyn cm$^{-2}$ in the polar region and $3 \times 10^{-3}$ dyn cm$^{-2}$ above the quiet Sun at the height of $7 \times10^4$ km. The boundary of the polar coronal hole and the quiet Sun must be determined by the total pressure balance between the coronal hole and the corona in the quiet Sun, and the polar corona expands to the lower latitude until it reaches the magnetic plus plasma pressure equilibrium with the quiet Sun corona. This may be the reason why the fast solar wind from the polar region reaches the lower latitude \citep{McCo00}.   

\subsection{Polar coronal activities}
\citet{Defor97} investigated the correlation between solar plumes and polar magnetic field with {\it SOHO}, and found that there is opposite magnetic flux close to the foot-point of solar plumes. In the polar coronal hole, X-ray jets are observed with high occurrence rate (60 polar X-ray jets per day on average) \citep{Kamio07, Cirtain07,Sav07}. \citet{Shimojo09} revealed that the opposite magnetic fields close to the kG-patches are related to the occurrence of the X-ray jets. Distribution of such opposite magnetic fields around a kG-patch is indeed seen in the (signed) magnetic field map in the North polar region (Fig. \ref{npole3}), and may be responsible for various coronal activities. 

\acknowledgments
The MILOS code was developed by Orozco Su\'arez. The authors express sincere appreciation to his work, and thank him for allowing us to use the software. We acknowledge the useful comments and encouragements by D. Orozco Su\'arez, L. Harra, E. Hiei. R. Ishikawa, R. Kano, N. Narukage, K. D. Leka, and J. Okamoto. Hinode is a Japanese mission developed and launched by ISAS/JAXA, collaborating with NAOJ as a domestic partner, NASA and STFC (UK) as international partners. Scientific operation of the Hinode mission is conducted by the Hinode science team organized at ISAS/JAXA. This team mainly consists of scientists from institutes in the partner countries. Support for the post-launch operation is provided by JAXA and NAOJ (Japan), STFC (U.K.), NASA, ESA, and NSC (Norway).


\begin{thebibliography}{}
\bibitem[Babcock and Babcock (1955)]{Babcock55}
	Babcock, H. W., \& Babcock, H. D. 1955, \apj, 121, 349
\bibitem[Bellot Rubio (2009)]{Bellot09}
	Bellot Rubio, L. R. 2009, \apj, 700, 284
\bibitem[Benevolenskaya (2004)]{Bene04}
	Benevolenskaya, E. E. 2004, A\&A, 428, L5
\bibitem[Blanco Rodriguez et al.(2007)]{Blanco07}
	Blanco Rodriguez, J., Okunev, O. V., Puschmann, K. G., Kneer,
	F., \& S\'anchez Andrade Nu\~no, B. 2007,
	A\&A, 474, 251
\bibitem[Cirtain et al.(2007)]{Cirtain07}
	Cirtain, J. W., et al. 2007, Science, 318, 1580
\bibitem[DeForest et al. (1997)]{Defor97}
	DeForest, C. E., et al. 1997, \solphys, 175, 393
\bibitem[Fox et al. (1998)]{Fox98} 
	Fox, P., McIntosh, P. \& Wlison, P., R. 1989, \solphys, 120, 285
\bibitem[Golub et al. (2007)]{Golub07}
	Golub, L., et al. 2007, \solphys, 243, 63
\bibitem[Fujimura and Tsuneta (2009)]{fuji}
	Fujimura, D., \& Tsuneta, S. 2009, \apj, 702, 1443 
\bibitem[Homann et al.(1997)]{Homann97}
	Homann, T., Kneer, F., \& Makarov, V. I. 1997, \solphys, 175, 81
\bibitem[Ichimoto et al.(2008)]{Ichi08}
	Ichimoto, K., et al. 2008, \solphys, 249, 233
\bibitem[Ishikawa and Tsuneta (2009)]{ishi09}
	Ishikawa, R., \& Tsuneta, S., 2009, A\&A, 636
\bibitem[Ishikawa, Tsuneta, \& Jur\v{c}\'{a}k (2010)]{ishi10}
    Ishikawa, R., Tsuneta, S., \& Jur\v{c}\'{a}k, J., 2010, \apj, 713, 1310 
\bibitem[Kamio et al.(2007)]{Kamio07}
	Kamio, S., et al. 2007, \pasj, 59, S757	
\bibitem[Kano et al.(2008a)]{Kano08a}
	Kano, R., et al. 2008a, \solphys, 249, 263
\bibitem[Kano et al.(2008b)]{Kano08b}
	Kano, R., et al. 2008b, \pasj, 60, S827	
\bibitem[Kojima et al.(2001)]{Kojima01}
	Kojima, M., et al. 2001, \jgr, 106, 15677
\bibitem[Kosugi et al.(2007)]{Kosugi07}
	Kosugi, T., et al. 2007, \solphys, 243, 3
\bibitem[Lin et al.(1994)]{Lin94}
	Lin, H., Virski, J., \& Zirin, H. 1994, \solphys, 155, 243
\bibitem[Lites (1996)]{Lites96}
	Lites, B. W. 1996, \solphys, 163, 223
\bibitem[Lites et al.(2001)]{Lites01}
	Lites, B. W., Elmore, D. F., \& Streander, K. V. 2001, ASP Conf. Ser. 236,
    Advanced Solar Polarimetry-Theory, Observation, and Instrumentation,
    ed. M. Sigwarth (San Francisco, CA: ASP), 33
\bibitem[Lites et al.(2008)]{Lites08}
    Lites, B. W., et al. 2008, \apj, 672, 1273 		
\bibitem[McComas et al.(2000)]{McCo00}
	McComas, D. J., et al. 2000, \jgr, 105, 10419
\bibitem[Okunev and Kneer (2004)]{Oku04}
	Okunev, O. V., \& Kneer, F. 2004, A\&A, 425, 321
\bibitem[Okunev and Kneer (2005)]{Oku05}
	Okunev, O. V., \& Kneer, F. 2005, A\&A, 439, 323
\bibitem[Orozco Su\'arez and del Toro Iniesta (2007)]{david07}
	Orozco Su\'arez, D., \& del Toro Iniesta J. C. 2007, A\&A,
		462, 1137 
\bibitem[Orozco Su\'arez et al.(2007a)]{david07a}
	Orozco Su\'arez, D., et al. 2007a, \apj, 670, L61
\bibitem[Orozco Su\'arez et al. (2007b)]{david07b}
	Orozco Su\'arez, D., et al. 2007b, \pasj, 59, S837	
\bibitem[Savcheva et al. (2007)]{Sav07}
	Savcheva, A., et al. 2007, \pasj, 59, S771	
\bibitem[Severny (1971)]{Severny71}
	Severny, A., B. 1971, Q. Jl R. astr. Soc., 12, 363--379
\bibitem[Shimizu et al. (2008)]{Shimizu08}
	Shimizu, T., et al. 2008, \solphys, 249, 221
\bibitem[Shimojo \& Tsuneta (2009)]{Shimojo09}
	Shimojo, M., \& Tsuneta, S., 2009, \apj, 706, L145
\bibitem[Shiota  (2009)]{Shiota09}
	Shiota, D. 2009, private communication
\bibitem[Svalgaard et al. (1978)]{svalgaard78}
    Svalgaard, L., Duvall, Jr, T. L., \& Scherrer, P. H., 1978, \solphys, 58, 225 
\bibitem[Suematsu et al.(2008)]{Suematsu08}
	Suematsu, Y., et al. 2008, \solphys, 249, 197
\bibitem[Suzuki and Inutsuka (2006)]{Suzuki06}
	Suzuki, T. K. \& Inutsuka, S. 2006, \jgr, 111, 6101
\bibitem[Tang and Wang (1991)]{Tang91}
	Tang, F. \& Wang, H. 1991, \solphys, 132, 247
\bibitem[Tsuneta et al.(2008a)]{tsuneta08}
	Tsuneta, S., et al. 2008a, \apj, 688, L1374
\bibitem[Tsuneta et al.(2008b)]{tsuneta08a}
	Tsuneta, S., et al. 2008b, \solphys, 249, 167
\bibitem[Tu et al.(2005)]{Tu05}
	Tu, C.-Y., Zhou, C., Marsch, E., Xia, L.-D., Zhao, L., Wang,
		J.-X., \& Wilhelm, K. 2005, Science, 308, 519 
\bibitem[Wang et al. (1989)]{Wang89}
	Wang, Y.-M., Nash, A. G., \& Sheeley, N. R., Jr. 1989, \apj, 347, 529
\bibitem[Wang et al. (1990)]{Wang90}
	Wang, Y.-M., \& Sheeley, N. R., Jr. 1990, \apj, 355, 726
\end{thebibliography}
\end{document}